\documentstyle[epsf]{article}
\def\be{\begin{equation}}
\def\ee{\end{equation}}
\def\bea{\begin{eqnarray}}
\def\eea{\end{eqnarray}}
\def\l{\label}
\def\ct{\cite}
\def\r{\ref}

\def\Th{\Theta}

\def\udot{\dot{u}}
\def\4pig{\sfrac{4\pi G}{c^{4}}}

\def\eps{\epsilon}

\def\D{\mbox{D}}

\def\hsp5{\hspace{5mm}}
\newcommand{\sfrac}[2]{{\textstyle{#1\over#2}}}
\def\case#1/#2{\textstyle\frac{#1}{#2}}
\def\cqg{{\em Class. Quantum Grav.} }
\def\grg{{\em Gen. Rel. Grav.} }
\def\prd{{\em Phys. Rev.} D }

\newcommand{\enl}{\\\hfill\rule{0pt}{0pt}} 
\title{\sc Deviation of geodesics in FLRW spacetime geometries}
\author{{\sc George F R Ellis\thanks{e-mail: ellis@maths.uct.ac.za}
\ \& Henk van Elst\thanks{e-mail: henk@gmunu.mth.uct.ac.za}}\\
{\small\em Department of Mathematics and Applied Mathematics,
University of Cape Town, Rondebosch 7700}\\
{\small\em Cape Town, South Africa}}
\date{\normalsize{August 29, 1997}}
\begin{document}
\sloppy
\maketitle
\begin{abstract}
The geodesic deviation equation (`GDE') provides an elegant tool to
investigate the timelike, null and spacelike structure of spacetime
geometries. Here we employ the GDE to review these structures
within the Friedmann--Lema\^{\i}tre--Robertson--Walker (`FLRW')
models, where we assume the sources to be given by a
non-interacting mixture of incoherent matter and radiation, and we
also take a non-zero cosmological constant into account. For each
causal case we present examples of solutions to the GDE and we
discuss the interpretation of the related first integrals. The de
Sitter spacetime geometry is treated separately.

\end{abstract}
\vspace{10cm}
\begin{flushleft}
{\em This paper is dedicated to Engelbert Sch\"{u}cking}
\end{flushleft}
\newpage
\section{Introduction}
It has been known for a long while that the geodesic deviation
equation (`GDE'), first obtained by J L Synge \ct{syn34,syn60},
provides a very elegant way of understanding features of curved
spaces, and, as pointed out by Pirani \ct{pir56,pir57}, gives an
invariant way of characterising the nature of gravitational forces
in spacetime.  As such, it is a useful tool to use in examining
specific exact solutions of the Einstein field equations (`EFE').
Indeed, it may be claimed that the GDE is one of the most important
equations in relativity, as this is {\em how\/} one measures
spacetime curvature\footnote{And so is analogous to the Lorentz
force law in electrodynamics; cf. Misner, Thorne and Wheeler, Ch.3,
Box 3.1 \ct{mtw73}.}. This latter aspect has been discussed in some
depth by Szekeres \ct{sze65}.\enl

The GDE determines the second rate of change of the deviation
vectors for a congruence of geodesics of arbitrary causal
character, i.e., their relative acceleration. Consider the
normalised tangent vector field $V^{a}$ for such a congruence,
parametrised by an affine parameter $v$. Then
\be
\l{geod}
V^{a} := \frac{dx^{a}(v)}{dv}\ , \hsp5 V_{a}\,V^{a} := \eps \ ,
\hsp5 0 = {\delta V^{a} \over \delta v}= V^{b}\nabla_{b}V^{a} 
\ ,
\ee
where $\eps = +\,1, \,0, \,-\,1$ if the geodesics are spacelike,
null, or timelike, respectively, and we define covariant
derivativion {\em along\/} the geodesics by $\delta T^{a..}{}_{b..}
/\delta v := V^{c}\nabla_{c}T^{a..}{}_{b..}$ for any tensor
$T^{a..}{}_{b..}$. A deviation vector $\eta^{a} := dx^{a}(w)/dw$
for the congruence, which can be thought of as linking pairs of
neighbouring geodesics in the congruence, commutes\footnote{The Lie
derivative of $\eta^{a}$ along the integral curves of $V^{a}$ is
zero; see, e.g., Schouten \ct{sch54}.} with $V^{a}$, so
\be
\l{com}
{\delta\eta^{a} \over \delta v} = 
\eta^{b}\nabla_{b}V^{a} \ .
\ee
It follows that their scalar product is constant along the
geodesics:
\be
\l{vetaang}
{\delta  (\eta_{a}V^{a}) \over \delta v}  = 0 \hsp5
\Leftrightarrow \hsp5 (\eta_{a}V^{a}) = \mbox{const} \ .
\ee
To simplify the relevant equations, we always choose them
orthogonal:
\be
\l{orth}
\eta_{a}\,V^{a} = 0 \ .
\ee
The general GDE takes the form 
\be
\l{gde}
\frac{\delta^{2}\eta^{a}}{\delta v^{2}}
= - \,R^{a}\!_{bcd}\, V^{b}\,\eta^{c}\,V^{d} \ ,
\ee
(\,see, e.g., Synge and Schild \ct{synsch49}, Schouten \ct{sch54},
or Wald \ct{wal84}\,). The general solution to this second-order
differential equation along any geodesic $\gamma$ will have two
arbitrary constants (corresponding to the different congruences of
geodesics that might have $\gamma$ as a member).  There is a {\em
first integral\/} along any geodesic that relates the connecting
vectors for two {\em different\/} congruences which have one
central geodesic curve (with affine parameter $v$) in common. This
is
\be
\l{gdefi}
\eta_{1}{}_{a}\,{\delta\eta_{2} \over \delta v}^{a}
- \eta_{2}{}_{a}\,{\delta\eta_1 \over \delta v}^{a} = \mbox{const}
\ ,
\ee
and is completely independent of the curvature of the spacetime
manifold.\enl

The aim of this paper is to systematically use the GDE to explore
the geometry of the standard
Friedmann--Lema\^{\i}tre--Robertson--Walker (`FLRW') models of
relativistic cosmology (\,see, e.g.,
Refs. \ct{rob33,wei72,ell87}\,), solving the GDE for timelike, null
and spacelike geodesic congruences in these geometries; hence,
obtaining the Raychaudhuri equation \ct{ray55} determining the time
evolution of these models \ct{ehl61,ell71}, the Mattig
observational relations \ct{mat58} underlying the interpretation of
cosmological data \ct{san61}, and determining the nature of their
spatial 3-geometry \ct{rob33,ell87}. Also, we identify in each
case the first integral for the GDE and comment on its meaning, in
the null case leading to the usual reciprocity theorem \ct{ell71},
and in the timelike case obtaining generic solutions of the GDE via
this integral. Thus, our purpose is to characterise the major
geometrical and physical features of these spacetimes by use of the
GDE, hence showing the utility of this equation in obtaining all
the essential geometrical and dynamical results of standard
cosmology in a unified way. It is a pleasure to dedicate this paper
to Engelbert Sch\"{u}cking, who has made a major contribution to
obtaining clarity and elegance in understanding many features of
relativistic cosmology.

\subsection{The cosmological context}
In the cosmological situation we consider, we assume the sources of
the gravitational field to be a non-interacting mixture of
incoherent matter and radiation, to each of which the
phenomenological fluid description applies (\,see, e.g.,
Refs. \ct{ehl61} and \ct{ell71}\,). For completeness we also
include a cosmological constant $\Lambda$.

Notation used is as follows: $u^{a}$ is the normalised timelike
tangent vector field ($u_{a}\,u^{a} = -\,1$) to the fundamental
matter fluid flow, which is geodesic: $0 = u^{b}\nabla_{b}u^{a} :=
\udot^{a}$. The integral curves of $u^{a}$ are parameterised by the
proper time $t$ of comoving fundamental observers. We use standard
FLRW comoving coordinates:
\bea
\l{ds2}
ds^{2} & = & -\,dt^{2} + a^{2}(t)\,f_{\mu\nu}(x^{\rho})\,dx^{\mu}\,
dx^{\nu} \ , \hsp5
f_{\mu\nu}\,dx^{\mu}\,dx^{\nu} \  = \ dr^{2} + f^{2}(r)\,
(\,d\theta^{2} + \sin^{2}\theta\,d\phi^{2}\,) \ , \\
u^{a} & = & (\partial_{t})^{a} \ = \ \delta^{a}\!_{0} \ ,
\eea
where $a(t)$ denotes the time dependent scale factor, and the
function\footnote{Determined later by use of the 3-D spatial GDE.}
$f(r)$ relates to the intrinsic curvature of the spacelike
3-surfaces $\{t=\mbox{const}\}$ orthogonal to $u^{a}$. By spatial
homogeneity and isotropy, the covariant derivative of $u^{a}$
\ct{ehl61} reduces to
\be
\l{uder}
\nabla_{a}u_{b} = \sfrac{1}{3}\,\Th\,h_{ab} \ , \hsp5
\Th := \D_{a}u^{a} = 3\,\frac{\dot{a}}{a} \ .
\ee 
Here, $h_{ab}$ is the standard orthogonal projection tensor
\be
h_{ab} = g_{ab} + u_{a}\,u_{b} \hsp5 
\Rightarrow \hsp5 h_{\alpha\beta} = g_{\alpha\beta} \ ,
\ee
$\Th$ is the fluid rate of expansion, and the spatial derivative
operator (projected orthogonal to $u^{a}$ on all indices) is
denoted by $\D_{a}$ (\,cf. Ref. \ct{maa97}\,). It is a well-known
consequence of Eq. (\r{uder}) that FLRW spacetime geometries have
vanishing Weyl curvature (\,cf. Refs. \ct{ehl61} and \ct{ell71}\,),
\be
\l{zeroweyl}
C_{abcd} = 0 \ ;
\ee
the fluid matter flow neither generates tidal gravitational fields
nor causes propagation of gravitational waves.

\section{The Riemann curvature tensor}
\l{sec:riem}
In order to determine the explicit form of the GDE (\r{gde}), we
need the Riemann curvature tensor $R_{abcd}$. Because of
Eq. (\r{zeroweyl}), $R_{abcd}$ can be expressed purely in terms of
the Ricci curvature tensor $R_{ab}$, its trace $R$, and the metric:
\be
\l{riem}
R_{abcd} = \sfrac{1}{2}\,(\,R_{ac}\,g_{bd}-R_{ad}\,g_{bc}
+R_{bd}\,g_{ac}-R_{bc}\,g_{ad}\,)
- \sfrac{1}{6}\,R\,(\,g_{ac}\,g_{bd}-g_{ad}\,g_{bc}\,) \ .
\ee
The EFE algebraically determine $R_{ab}$ from the matter tensor
$T_{ab}$:\footnote{Geometrised units, characterised by $c = 1 =
8\pi G/c^{2}$, are used throughout.}
\be
\l{efe}
R_{ab} = T_{ab} - \sfrac{1}{2}\,T\,g_{ab} + \Lambda\,g_{ab}
\hsp5 \Rightarrow \hsp5 R = -\,T + 4\,\Lambda \ .
\ee
When the matter takes a `perfect fluid' form: 
\be
T_{ab} = (\mu+p)\,u_{a}\,u_{b} + p\,g_{ab}
\hsp5 \Rightarrow \hsp5 T = -\,(\mu-3p) \ ,
\ee
($\mu$ is the total energy density and $p$ the isotropic pressure),
the Ricci tensor expression is
\be
R_{ab} = (\mu+p)\,u_{a}\,u_{b} + \sfrac{1}{2}\,(\mu-p+2\Lambda)
\,g_{ab} \hsp5 \Rightarrow \hsp5 R = (\mu-3p) + 4\,\Lambda \ .
\ee
Thus, from Eq. (\r{riem}), the curvature tensor takes the form
\bea
\l{flrwrie}
R_{abcd} & = & \sfrac{1}{3}\,(\mu+\Lambda)\,(\,g_{ac}\,g_{bd}
-g_{ad}\,g_{bc}\,) \nonumber \\
& & \hsp5 + \ \sfrac{1}{2}\,(\mu+p)\,(\,g_{ac}\,u_{b}\,u_{d}
-g_{ad}\,u_{b}\,u_{c}+g_{bd}\,u_{a}\,u_{c}-g_{bc}\,u_{a}\,u_{d}\,)\,.
\eea
Then, for any normalised vector field $V^{a}$: $V_{a}\,V^{a} =
\eps$, by a straightforward contraction one obtains from
Eq. (\r{flrwrie}) the source term in the GDE:
\bea
\l{conrie}
R_{abcd}\,V^{b}\,V^{d} & = & \sfrac{1}{3}\,(\mu+\Lambda)\,
(\,\eps\,g_{ac}- V_{a}\,V_{c}\,) \nonumber \\
& & \hsp5 + \ \sfrac{1}{2}\,(\mu+p)\,[\ (V_{b}u^{b})^{2}\,g_{ac}
- 2\,(V_{b}u^{b})\,u_{(a}\,V_{c)} + \eps\,u_{a}\,u_{c}\ ] \ .
\eea
We will also want the GDE in the spacelike 3-surfaces
$\{t=\mbox{const}\}$ orthogonal to $u^{a}$, which are 3-spaces of
maximal symmetry. In the FLRW case, the Gau\ss\ embedding equation
provides the relation
\be
\l{3curv}
{}^{3}\!R_{abcd}= (R_{abcd})_{\perp} - \sfrac{1}{9}\,\Th^{2}\,
(\,h_{ac}\,h_{bd}-h_{ad}\,h_{bc}\,)
\ee
for the 3-D Riemann curvature. From Eq. (\r{flrwrie}), which made
use of the EFE, one has
\be
(R_{abcd})_{\perp} = \sfrac{1}{3}\,(\mu+\Lambda)\,
(\,h_{ac}\,h_{bd}-h_{ad}\,h_{bc}\,) \ ,
\ee
so that Eq. (\r{3curv}) becomes
\be
\l{cc}
{}^{3}\!R_{abcd} = K(t)\,(\,h_{ac}\,h_{bd}-h_{ad}\,h_{bc}\,) \ ,
\ee
where the spatial curvature scalar $K(t)$ is given by 
\be
\l{eq:gauss}
K(t) := \sfrac{1}{6}\,{}^{3}\!R = \sfrac{1}{3}\,(\,\mu
-\sfrac{1}{3}\,\Th^{2}+\Lambda\,) \ .
\ee
This factor will determine the 3-D spatial GDE\footnote{See section
\r{sec:gengeo} below.} source term:
\be
\l{3d}
{}^{3}\!R_{abcd}\,V^{b}\,V^{d} = K\,(\,h_{ac}-V_{a}\,V_{c}\,) \ ,
\ee
where $V_{a}\,V^{a}=1$ and $V_{a}\,u^{a}=0$.

\section{The geodesics}
\l{sec:geo}
Before turning to address the GDE, we need to solve for the
geodesic curves along which the GDE will be integrated. Now the
fundamental 4-velocity $u^{a}=\delta^{a}\!_{0}$ is a geodesic
vector field. Any other geodesic can be transformed to have a
purely {\em radial\/} spatial part by suitable choice of local
coordinates (because the FLRW geometry is isotropic about every
point). Hence, w.o.l.g., radial geodesics are considered, with the
origin of the local coordinates $r = 0$ at the starting point
$v=0$, so that in all cases we will have $x^{2} = \theta =
\mbox{const}$, $x^{3} = \phi = \mbox{const}$ $\Rightarrow 0 = V^{2}
= V^{3}$.

It is convenient to decompose a general geodesic tangent vector
field $V^{a}$ into parts parallel and orthogonal to $u^{a}$:
\be
\l{vdec}
V^{a} := E\,u^{a} + P\,e^{a} \ ,
\ee
where $e^{a} = a^{-1}\,(\partial_{r})^{a} =
a^{-1}\,\delta^{a}\!_{1}$, $e_{a}\,e^{a} = 1$, $e_{a}\,u^{a} = 0$,
such that
\be
V_{a}\,V^{a} = \eps = -\,E^{2} + P^{2} \ , \hsp5
-(V_{a}u^{a}) = E \ , \hsp5 P = (\eps+E^{2})^{1/2} \ .
\ee
As $e^{a}$ spans a radial direction, $P \geq 0$. By spatial
homogeneity and isotropy\footnote{Even though $e^{a}$ is {\em
not\/} invariantly defined, the $1+3$ covariant discussion of LRS
perfect fluid spacetime geometries given in Ref. \ct{hveell96}
still applies. As such, $e^{a}$ is the Fermi-transported (along
$u^{a}$) unit tangent of a geodesic and shearfree spacelike
congruence. Furthermore, in the given context also its spatial
rotation vanishes.}, for a congruence of radial normalised
geodesics, starting off isotropically from $r = 0 \Leftrightarrow v
= 0$ (so $E\,|_{v=0} = \mbox{const}$ for all of them),
\be
\l{sh}
0 = \D_{a}E = \D_{a}P \ .
\ee
To determine $E$, note that
\bea
\l{eevol}
-\,\frac{\delta(V_{a}u^{a})}{\delta v}
& = & -\,V^{b}\nabla_{b}(V_{a}u^{a})
\ = \ -\,V^{a}\,(\nabla_{b}u_{a})\,V^{b} \nonumber \\
& = & -\,\sfrac{1}{3}\,\Th\,h_{ab}\,V^{a}\,V^{b}
\ = \ -\,\sfrac{1}{3}\,\Th\,[\ \eps + (V_{a}u^{a})^{2}\ ] \ .
\eea
Thus, we need to solve
\be
\frac{dt}{dv} = V^{0} = E = -(V_{a}u^{a}) \ , \hsp5
\frac{1}{(\eps+E^{2})}\,\frac{dE}{dv} = -\,\frac{1}{a}\,
\frac{da}{dt} \ ;
\ee
so
\be
\frac{E}{(\eps+E^{2})}\,\frac{dE}{dv}  =  -\,\frac{1}{a}\,
\frac{da}{dt}\,\frac{dt}{dv} 
\hsp5 \Leftrightarrow \hsp5
\sfrac{1}{2}\,\frac{d}{dv}\,\ln
\left[\,\frac{(\eps+E^{2})}{(\eps+E_{0}^{2})}\,\right]
\ = \ \frac{d}{dv}\,\ln\left[\,\left(\frac{a}{a_{0}}\right)\,
\right]^{-1} \ .
\ee
Integrating, we obtain
\be
\frac{(\eps+E^{2})}{(\eps+E_{0}^{2})}
= \left(\frac{a_{0}}{a}\right)^{2} \ .
\ee
Now solving for $E$,
\be
E^{2} = (\eps+E_{0}^{2})\,\left(\frac{a_{0}}{a}\right)^{2} - \eps
\ ,
\ee
which implies
\be
\l{dtdv}
\frac{dt}{dv} = V^{0} = E(a) = \pm\,\left[\ (\eps+E_{0}^{2})\,
\left(\frac{a_{0}}{a}\right)^{2} - \eps\ \right]^{1/2} \ ,
\ee
with a ``$+$'' for {\em future\/}-directed vectors $V^{a}$ and a
``$-$'' for {\em past\/}-directed ones. Also, with Eq. (\r{ds2}),
\bea
V^{a}\,g_{ab}\,V^{b} = -\,(V^{0})^{2} + a^{2}\,(V^{1})^{2}
\ = \ \eps \hsp5 \Leftrightarrow \hsp5
V^{a}\,h_{ab}\,V^{b} = \eps + E^{2} = a^{2}\,(V^{1})^{2} \ ,
\eea
so
\be
\l{drdv}
\frac{dr}{dv} = V^{1} = \frac{P(a)}{a} = \left[\,\frac{\eps
+E^{2}(a)}{a^{2}}\,\right]^{1/2} \ ,
\ee
which, for later reference, can also be cast into the form
\be
\l{dldv}
d\ell := a\,dr
= (\eps+E_{0}^{2})^{1/2}\left(\frac{a_{0}}{a} \right)
dv \ ,
\ee
the definition coming from Eq. (\r{ds2}). Hence,
\be
\frac{dt}{dr} = \frac{dt/dv}{dr/dv}
= \pm\,\frac{E(a)}{[\,a^{-2}\,(\eps+E^{2}(a))\,]^{1/2}}
= \pm\,\frac{a^{2}\,E(a)}{a_{0}\,(\eps+E_{0}^{2})^{1/2}} \ ,
\ee
and so
\be
\frac{dt}{dr} = \pm\,a(t)\,\left[\ 1 - \eps\,
\left(\frac{a(t)}{\alpha_{0}}\right)^{2}\ \right]^{1/2} \ , \hsp5
\alpha_{0} := \pm\,a_{0}\,(\eps+E_{0}^{2})^{1/2} \ .
\ee
%

\subsection{Timelike}
For timelike vector fields, $\eps = -\,1$. If we have $V^{a}$ {\em
initially\/} parallel to $u^{a}$, then $E_{0}^{2} = 1$, and so
$dt/dv = 1$ and $dr/dv = 0$, confirming that $V^{a}$ then {\em
remains\/} parallel to $u^{a}$ (which is geodesic). Otherwise, for
future-directed timelike geodesics $V^{a}$ that have a {\em
non-zero\/} initial hyperbolic angle of tilt with $u^{a}$ (such
that $E_{0}^{2} > 1$), the following relations apply:
\bea
\l{tldtdv}
\frac{dt}{dv} = \left[\ 1 + (E_{0}^{2}-1)\,\left(
\frac{a_{0}}{a}\right)^{2}\ \right]^{1/2} \ , \hsp5
\frac{dr}{dv}  = (E_{0}^{2}-1)^{1/2}\,\left(
\frac{a_{0}}{a^{2}}\right) \ .
\eea
%

\subsection{Spacelike}
\l{sec:slgeod}
For spacelike vector fields, $\eps=+\,1$.  Setting $E_{0}=0$ means
starting off {\em orthogonally\/}, but these geodesics do {\em
not\/} remain orthogonal to the flow lines, and so do not remain
within the spacelike 3-surfaces $\{t = \mbox{const}\}$. Indeed,
from Eqs. (\r{eevol}) and (\r{dtdv})
\bea
& & -(V_{a}u^{a})\,|_{P} = 0 \ , \hsp5 \Th\,|_{P} > 0
\nonumber \\
& & \Rightarrow \ -(V_{a}u^{a}) = E < 0 \hsp5 \mbox{nearby}
\hsp5 \Rightarrow \ \frac{dt}{dv} < 0
\ , 
\eea
showing that the {\em geodesic\/}, nearby spacelike 3-surfaces bend
{\em down\/} (into the past) relative to the spacelike 3-surfaces
$\{t = \mbox{const}\}$. In this case $\alpha_{0} = \pm\,a_{0}$, and
$$
\frac{dt}{dv} = -\,\left[\ \left(\frac{a_{0}}{a}\right)^{2}
- 1\ \right]^{1/2} \ .
$$
So, with $dr/dv = (a_{0}/a^{2})$, we find
\be
\frac{dt}{dr} = -\,a\,\left[\ 1 - \left(\frac{a}{a_{0}}
\right)^{2}\ \right]^{1/2} \ .
\ee
The geodesic 3-surfaces give the best slicing of a spacetime in
order to approximate Newtonian theory in a general spacetime ---
see the discussion by Ehlers \ct{ehl73} --- and have been studied
in the FLRW context by Rindler \ct{rin81}, Page \ct{pag83}, and
Ellis and Matravers \ct{ellmat85}.

The simplest dynamical case is the spatially flat Einstein--de
Sitter model, which has (pressure-free) incoherent matter as a
source, and $\Lambda = 0$. Here, the length scale factor takes the
functional form $a(t) = a_{0}\, [\,\sfrac{3}{2}H_{0}\,t\,]^{2/3}$,
where $H_{0}$ is the value of the Hubble parameter $H :=
(1/a)\,(da/dt)$ at time $t_0$. Hence, we obtain (\,note that $t
\leq t_{0}$\,)
\be
r(t,t_{0})  = - \,\frac{1}{a_{0}\,(3H_{0}/2)^{2/3}}\,
\int_{t_{0}}^{t} \frac{dy}{y^{2/3}\,
[\ 1 - (3H_{0}/2)^{4/3}\,y^{4/3}\ ]^{1/2}} \ ,
\ee
leading to an elliptic integral which gives the value of $r$ at
time $t$, starting off orthogonally at $r = 0$ and time $t_{0} =
\sfrac{2}{3}\,H_{0}^{-1}$. 

\subsection{Null}
In the case of null congruences, $\eps=0$. Then it follows for the
{\em past\/}-directed case that
$$
\frac{dt}{dv} = -\,\left[\ (E_{0}^{2})\,\left(
\frac{a_{0}}{a}\right)^{2} \ \right]^{1/2} = - \,|E_{0}|
\left(\frac{a_{0}}{a}\right) \ ,
$$
and, as $dr/dv = |E_{0}|\,(a_{0}/a^{2})$,
\be
\frac{dt}{dr} = -\,a(t) \ .
\ee
Alternatively, we can use the fact that $\xi_{a} := a(t)\,u_{a}$ is
a conformal Killing vector field: $\nabla_{a}\xi_{b} =
\dot{a}(t)\,g_{ab}$. Thus, for any geodesic vector field $k^{a}$,
\be
k^{b}\nabla_{b}(\xi_{a}k^{a}) = \dot{a}(t)\,k^{a}\,g_{ab}\,k^{b}
= \dot{a}(t)\,k_{a}\,k^{a} \ ,
\ee
and in the particular case that $k^{a}$ is null:
\bea
k_{a}\,k^{a} & = & 0 \hsp5 \Rightarrow \hsp5
\xi_{a}\,k^{a} \ = \ a(t)\,
u_{a}\,k^{a} \ = \ \mbox{const} \nonumber \\
& \Rightarrow &
(k_{a}u^{a})(t) \ = \ \frac{\mbox{const}}{a(t)} \ .
\eea
Relating this to the redshift, $z$, defined by
\be
\l{red}
(1+z) := \frac{(k_{a}u^{a})_{e}}{(k_{b}u^{b})_{0}}
= \frac{E_{e}}{E_{0}} = \left(\frac{a_{0}}{a(t_{e})}\right) \ ,
\ee
for past-directed radial null geodesics it follows that
\be
\l{nullcomp}
k^{a} = \frac{a_{0}}{a(t)}\left(\,-1,\,\frac{1}{a(t)},\,0,\,0\,
\right)^{T} \ , \hsp5
\frac{dt}{dv} = k^{0} = E = -\,(1+z) \ ,
\ee
where we have set $E_{0} = -\,1$ by choice of the affine parameter
$v$.

\section{The geodesic deviation equation}
\subsection{The deviation vectors}
The basic equations relating the geodesic vector $V^{a}$ and
orthogonal deviation vector $\eta^a$ have been given above, see
Eqs. (\r{geod}) - (\r{gde}). We now restrict the deviation vector
further.

\subsubsection{The screen space}
When $V^{a}$ is {\em not\/} parallel to $u^{a}$, the vector
$\eta^{a}$ lies in the {\em screen space\/} of $u^{a}$ (i.e., the
spacelike 2-surface orthogonal to both, $u^{a}$ and $V^{a}$) iff,
additionally to $(\eta_{a}V^{a}) = 0$, $\eta^{a}$ also lies in the
rest 3-space of $u^{a}$, i.e., $(\eta_{a}u^{a}) = 0$. We can choose
this to be true {\em initially\/}; will it be maintained along the
integral curves of any geodesic vector field $V^{a}$?  With
Eq. (\r{uder}), we have
\bea
{\delta (\eta_{a}u^{a}) \over \delta v}
= \sfrac{1}{3}\,\Th\,h_{ab}\,\eta^{a}\,V^{b}
+ u_{a}\,\eta^{b}\nabla_{b}V^{a}
= \sfrac{2}{3}\,\Th\,h_{ab}\,\eta^{[a}\,V^{b]}
+ \eta^{b}\nabla_{b}(V_{a}u^{a}) \ ,
\eea
so
\be
\l{c1}
{\delta (\eta_{a}u^{a})\over \delta v}
= \eta^{b}\nabla_{b}(V_{a}u^{a}) \ ,
\ee
which will be zero, if $\eta^{b}\nabla_{b}(V_{a}u^{a}) = 0$, and
this will be true for the congruences we consider
(\,cf. Eq. (\r{sh})\,). Propagation of condition (\r{c1}) along the
integral curves of $u^{a}$ then confirms its preservation. This can
be seen as follows. The fact that $V^{a}$ and $\eta^{a}$ commute,
Eq. (\r{com}), gives rise to the relation
\bea
0 = u_{a}\,[\ u^{c}(\nabla_{c}V^{b})\,(\nabla_{b}\eta^{a})
+ V^{b}\,u^{c}\nabla_{c}\nabla_{b}\eta^{a}
- u^{c}(\nabla_{c}\eta^{b})\,(\nabla_{b}V^{a})
- \eta^{b}\,u^{c}\nabla_{c}\nabla_{b}V^{a}\ ] \ ,
\eea
which is used to eliminate the respective terms in the
``dot''-derivative of condition (\r{c1}). Hence, with
Eq. (\r{uder}),
\be
[\ V^{b}\nabla_{b}(\eta_{a}u^{a}) - \eta^{b}\nabla_{b}
(V_{a}u^{a})\ ]\,\dot{}
= \sfrac{2}{3}\,h_{ab}\,[\ \Th\,\eta^{[a}\,V^{b]}\ ]\,\dot{} = 0
\ , 
\ee
which vanishes because $h_{ab}$ is symmetric in its indices. So,
the consistent solution to these equations is
\be
\l{orthog}
(\eta_{a}u^{a}) = 0 \ , \hsp5
(\eta_{a}V^{a}) = 0 \ , \hsp5
\D_{a}(V_{b}u^{b}) = 0 \ ;
\ee
i.e., $\eta^{a}$ starts and remains within the rest 3-spaces of
$u^{a}$, and it also remains orthogonal to $V^{a}$, which has a
constant scalar product with $u^{a}$ in these rest 3-spaces. From
now on we will assume these relations hold.

\subsubsection{The force term}
The ``force term'' (\,cf., e.g., Pirani \ct{pir56}\,) for the
general GDE (\r{gde}) for geodesic congruences of either timelike,
null or spacelike causal character, specialised to the FLRW case,
can now be evaluated from Eqs. (\r{conrie}) and (\r{orthog}) to
yield
\be
\l{devfor2}
R^{a}\!_{bcd}\,V^{b}\,\eta^{c}\,V^{d}  = [\ \eps\,\sfrac{1}{3}\,
(\mu+\Lambda)\, + \sfrac{1}{2}\,(\mu+p)\,E^{2}\ ]\ \eta^{a}
\ee
where, as before, $-(V_{a}u^{a}) = E$. Note that this force term is
proportional to $\eta^{a}$ itself, i.e., according to the GDE
(\r{gde}) only the magnitude $\eta$ will change along a geodesic,
while its spatial orientation will remain fixed\footnote{Also,
Eq. (\r{devfor2}) has {\em no\/} component proportional to $u^{a}$,
confirming the consistency of the above screen space
analysis.}. Consequently, the GDE (\r{gde}) reduces to give just a
{\em single\/} differential relation for the scalar quantity
$\eta$.  This reflects the spatial isotropy of the Riemann
curvature tensor about every point in the present situation;
anisotropic effects as induced, e.g., by non-zero electric Weyl
curvature, $E_{ab}$, or shear viscosity, $\pi_{ab}$, are not
involved.\enl

We deal, now, with three cases: the GDE for a fundamental observer,
for past-directed geodesic null congruences, and for other families
of geodesics.

\subsection{Geodesic deviation for a fundamental observer}
{\bf Case 1:} $V^{a} = u^{a}$ for the {\em central\/} geodesic. In
this case the affine parameter coincides with the proper time of
the central fundamental observer, i.e., $v=t$. From
Eq. (\r{devfor2}), with $\eps = -\,1$ and $E=1$,
\be
\l{tlfor}
R_{abcd}\,u^{b}\,\eta^{c}\,u^{d} = \sfrac{1}{6}\,(\mu+3p)\,\eta_{a}
- \sfrac{1}{3}\,\Lambda\,\eta_{a} \ .
\ee
Let the deviation vector be $\eta^{a}=\ell\,e^{a}$,
$e_{a}\,e^{a}=1$, $e_{a}\,u^{a}=0$, such that it connects
neighbouring flow lines in the radial direction. Then $\delta
e^{a}/\delta t = u^{b}\nabla_{b}e^{a}=0$ (as there is no shear or
vorticity!), i.e., a basis is used which is parallelly propagated
along $u^{a}$, and Eq. (\r{gde}) gives
\be
\l{tlgde}
\frac{d^{2}\ell}{dt^{2}} = -\,\sfrac{1}{6}\,(\mu+3p)\,\ell
+ \sfrac{1}{3}\,\Lambda\,\ell \ ,
\ee
which is the Raychaudhuri equation \ct{ray55}. However, this
equation applies to {\em both\/} comoving matter of active
gravitational mass density $(\mu+3p)$, and to test matter that is
not comoving. On the basis of this relation, it is clear that for
positive active gravitational mass density and non-negative
cosmological constant\footnote{If $\Lambda < 0$, for $(\mu+3p) > 0$
there will be focusing anyway.} all families of past- and
future-directed timelike geodesics will experience focusing,
provided $(\mu+3p) > 2\,\Lambda$, and so gives rise to the standard
singularity theorems (\,see, e.g.,
Refs. \ct{ray55,ehl61,hawpen70,ell71}\,).

\subsubsection{Comoving matter}
For comoving matter, $V^{a} = u^{a} \Rightarrow |\,E_{0}\,| = 1
\Rightarrow |\,E\,| = 1$ for the {\em whole\/} family of
geodesics. Then, set $\ell = a$ and multiply by $da/dt$ to get
\be
0 = \frac{da}{dt}\,\frac{d^{2}a}{dt^{2}}
+ \sfrac{1}{6}\,(\mu+3p)\,a\,\frac{da}{dt} - \sfrac{1}{3}\,
\Lambda\,a\,\frac{da}{dt} \ .
\ee
Using the conservation equation for comoving matter,
\be
\frac{d\mu}{dt} = -\,\frac{3}{a}\,\frac{da}{dt}\,(\mu+p)
\hsp5 \Rightarrow \hsp5
\frac{d(\mu\,a^{2})}{dt} = - \,(\mu+3p)\,a\,\frac{da}{dt} \ ,
\ee
one finds the familiar Friedmann equation
\be
\left(\frac{da}{dt}\right)^{2} - \sfrac{1}{3}\,\mu\,a^{2}
- \sfrac{1}{3}\,\Lambda\,a^{2}
= -\,k \ , \hsp5 k = \mbox{const} \ ,
\ee
giving the usual time evolution of $a(t)$ for a given equation of
state. In terms of invariants,
\be
\l{fried}
\left(\frac{1}{a}\,\frac{da}{dt}\right)^{2}
- \sfrac{1}{3}\,\mu - \sfrac{1}{3}\,\Lambda
= -\,\frac{k}{a^{2}} \ ,
\ee
which is just the trace of the Gau\ss\ equation,
Eq. (\ref{eq:gauss}), if we identify
\be
\l{3curvscl}
K = \frac{k}{a^{2}}
\ee
as the constant curvature of the spacelike 3-surfaces $\{t =
\mbox{const}\}$. Hence, we recover the standard dynamical equations
for the FLRW models from the GDE. As usual, whenever $K$ is
non-zero, by rescaling $a(t)$ by a constant the dimensionless
quantity $k$ can be normalised to $\pm\,1$, which is then the
curvature of the 3-spaces of maximal symmetry with metric
$f_{\mu\nu}\,dx^{\mu}\,dx^{\nu}$ (\,cf. Eq. (\r{ds2})\,).\enl

If one considers a non-interacting mixture of both incoherent
matter and radiation, one has
\be
\l{source}
\mu = 3\,H_{0}^{2}\,\Omega_{m_{0}}\left(\frac{a_{0}}{a}
\right)^{3} + 3\,H_{0}^{2}\,\Omega_{r_{0}}\left(\frac{a_{0}}{a}
\right)^{4} \ , \hsp5
p = H_{0}^{2}\,\Omega_{r_{0}}\left(\frac{a_{0}}{a}\right)^{4} \ .
\ee
Then, evaluating Eq. (\r{fried}) at $t=t_{0}$ shows that
\be
H_{0}^{2} - \sfrac{1}{3}\,(\mu_{m_{0}}+\mu_{r_{0}})
- \sfrac{1}{3}\,\Lambda
= -\,\frac{k}{a_{0}^{2}} \hsp5
\Leftrightarrow \hsp5 H_{0}^{2}\,(\Omega_{m_{0}}+\Omega_{r_{0}}
+\Omega_{\Lambda_{0}}-1)
= K_{0} \ ,
\ee
where
\be
K_{0} := \frac{k}{a_{0}^{2}} \ ,
\ee
and, as familiar, $\Omega_{i_{0}}$ denotes dimensionless
cosmological density parameters $\Omega_{i} := \mu_{i}/(3H^{2})$ at
$t=t_{0}$; $\Omega_{\Lambda} := \Lambda/(3H^{2})$ defines an
analogous quantity for the cosmological constant. Similarly,
evaluating the Raychaudhuri equation (\r{tlgde}) at $t=t_{0}$ gives
\be
q_{0} = - \,\frac{1}{3H_{0}^{2}}\left.\left(\frac{\ddot{a}}{a}
\right)\right|_{t_{0}} = \sfrac{1}{2}\,(\,\Omega_{m_{0}}
+ 2\,\Omega_{r_{0}} - 2\,\Omega_{\Lambda_{0}}\,)
\simeq \sfrac{1}{2}\,\Omega_{m_{0}} - \Omega_{\Lambda_{0}} \ ,
\ee
the $t_{0}$ value of the dimensionless cosmological deceleration
parameter $q := -\,(a\,d^{2}a/dt^{2})/(da/dt)^{2}$. These results
will be useful in deriving the observational relations for null
data (see section \r{sec:pastnull}).

\subsubsection{Non-comoving matter}
For isotropically distributed test matter moving with {\em other\/}
4-velocities about the fundamental observers, i.e., $V^{a}=v^{a}
\Rightarrow |\,E_{0}\,| > 1$, except for the {\em central\/} curve
of the congruence $v^{a}$ which coincides with $u^{a}$, we need to
obtain {\em other\/} solutions to the GDE for timelike curves,
evaluated along this central fundamental world line (where again
proper time $t$ is the same as the preferred affine parameter $v$,
and also here the deviation vectors have radial orientation). There
are two ways to do this.\enl

One way is to fully specify the matter source in the equations of
the previous discussion on the comoving matter case, solve these
equations to obtain the source term in the GDE (\ref{tlgde}), and
then solve the GDE to obtain its general solution (with two
arbitrary constants). In the case of the de Sitter universe, we
have $0 = \mu = p$, $\Lambda \neq 0$\footnote{Or, equivalently,
$(\mu+p) = 0$ $\Rightarrow$ $(\mu+3p) = -\,2\,\mu = \mbox{const}$,
giving an effective cosmological constant.}, so Eq. (\r{tlgde})
becomes
\be
\l{tlgde1}
0 = \frac{d^{2}\ell}{dt^{2}} - \sfrac{1}{3}\,\Lambda\,\ell
\ , 
\ee
and the solution is
\bea
\l{tldssol}
\ell(t) = \left\{ \begin{array}{ll}
          C_{1}\,\cosh(\alpha\,t) + C_{2}\,\sinh(\alpha\,t)
          & \Lambda > 0 \\
          C_{1}\,\cos(\alpha\,t) + C_{2}\,\sin(\alpha\,t)
          & \Lambda < 0 \\
                  \end{array} \right. \ ,
\eea
with $\alpha := (\sfrac{1}{3}\,|\,\Lambda\,|)^{1/2}$ and $C_{1}$
and $C_{2}$ integration constants carrying the dimension of
$\ell(t)$. This shows the deviation for {\em arbitrary\/} (i.e.,
independent of $|\,E_{0}\,| \geq 1$) timelike geodesics in the de
Sitter ($\Lambda > 0$) and anti-de Sitter ($\Lambda < 0$) cases.

When dynamical matter is present, life is more complex. Defining a
dimensionless conformal time variable $\tau$ by $dt/d\tau := a
\Rightarrow d^{2}t/d\tau^{2} = da/d\tau$, for a matter source
according to Eq. (\r{source}) the Friedmann equation (\r{fried})
yields
\be
\l{confried}
\frac{da}{d\tau} = [\ \sfrac{1}{3}\,\Lambda\,a^{4} - k\,a^{2}
+ a_{0}^{3}\,H_{0}^{2}\,\Omega_{m_{0}}a + a_{0}^{4}\,H_{0}^{2}\,
\Omega_{r_{0}}\ ]^{1/2} \ .
\ee
This can easily be solved when $\Lambda = 0$, for given value of
the spatial curvature parameter $k$. It follows that the GDE for
timelike congruences, Eq. (\r{tlgde}), can be rewritten as
\bea
\l{contlgde}
0 & = & \frac{d^{2}\ell}{d\tau^{2}} - \frac{1}{a}\,\frac{da}{d\tau}
\,\frac{d\ell}{d\tau} + \sfrac{1}{2}\,a_{0}^{2}\,H_{0}^{2}\,
[\ \Omega_{m_{0}}\left(\frac{a_{0}}{a}
\right) + 2\,\Omega_{r_{0}}\left(\frac{a_{0}}{a}
\right)^{2}\ ]\,\ell - \sfrac{1}{3}\,\Lambda\,a^{2}\,\ell \ ,
\eea
where $a=a(\tau)$, and $da/d\tau$ is determined through
Eq. (\r{confried}). Unfortunately, this linear homogeneous
second-order ordinary differential equation is very complicated,
except for the de Sitter universe (\,where $0 = \Omega_{m_0} =
\Omega_{r_0}$, $\Lambda \neq 0$\,), which we already considered.

To provide a simple example with dynamical matter, we fall back
onto the Einstein--de Sitter model, where $\Lambda=0$, $k=0$,
$\Omega_{r_{0}}=0 \Rightarrow \Omega_{m_{0}}=1$. In dimensionless
conformal time, the length scale factor is $a(\tau) =
\sfrac{1}{4}\,a_{0}^{3}\,H_{0}^{2}\,\tau^{2}$, and so the solution
to Eq. (\r{contlgde}), which then reduces to
\be
0 = \frac{d^{2}\ell}{d\tau^{2}} - \frac{2}{\tau}\,
\frac{d\ell}{d\tau} + \frac{2}{\tau^{2}}\,\ell \ ,
\ee
is given by
\be
\l{edsgdesol}
\ell(\tau) = C_{1}\,\tau + C_{2}\,\tau^{2} \ ;
\ee
again, the integration constants $C_{1}$ and $C_{2}$ carrying the
dimension of $\ell(\tau)$. Fixing initial conditions such as to
describe a set of test particles isotropically emanating from the
central reference geodesic at $\eta=\eta_{0}$, one has
$C_{2}=-\,C_{1}/\eta_{0}$.\enl

Another way to obtain solutions to the timelike GDE (\r{tlgde}) is
to use the first integral which relates {\em different\/} solutions
to the GDE along a central reference geodesic $\gamma_0$ (which is
common to both congruences, and on which the affine parameters
coincide and are equal to the preferred time coordinate, i.e.,
$v\,|_{\gamma_0}=t$).  Let $\eta_{1}$ relate to the fundamental
family of world lines and $\eta_{2}$ to another family. Then
$\eta_{1}=a(t)$, and as $dt/dv = -(v_{a}u^{a}) = E$ takes the value
$E=1$ on the central reference geodesic,
$(1/\eta_{1})\,(d\eta_{1}/dv)=H=\sfrac{1}{3}\,\Th$. Considering
parallel (radial) deviation vectors for the two families, we obtain
for their magnitudes
\be
\eta_{1}\,\frac{d\eta_{2}}{dt} - \eta_{2}\,\frac{d\eta_{1}}{dt}
= \mbox{const}
\hsp5 \Rightarrow \hsp5
\frac{d\eta_{2}}{dt} - \eta_{2}\,H(t) = \frac{\mbox{const}}{a(t)}
\ . 
\ee
In terms of initial data at time $t=t_{0}$,
\be
\left.\frac{d\eta_{2}}{dt}\,\right|_{t_{0}}
- \eta_{2}\,|_{t_{0}}\,H_{0} = \frac{\mbox{const}}{a_{0}} \ ,
\ee
which leads to
\be
\frac{d\eta_{2}}{dt} - \eta_{2}\,\frac{1}{a(t)}\,\frac{da(t)}{dt}
= \frac{a_{0}}{a(t)}\left[\ \left.\frac{d\eta_{2}}{dt}\,
\right|_{t_{0}} - \eta_{2}\,|_{t_{0}}\,H_{0}\ \right] \ ,
\ee
and so
\be
\frac{d}{dt}\left[\,\frac{\eta_{2}}{a(t)}\,\right]
= \frac{a_{0}}{a^{2}(t)}\left[\ \left.\frac{d\eta_{2}}{dt}\,
\right|_{t_{0}} - \eta_{2}\,|_{t_{0}}\,H_{0}\ \right] \ .
\ee
Then, integration yields
\be
\l{tlfisol}
\eta_{2}(t) = \eta_{2}\,|_{t_{0}}\left(\frac{a(t)}{a_{0}}\right)
+ \left[\ \left.\frac{d\eta_{2}}{dt}\,\right|_{t_{0}}
- \eta_{2}\,|_{t_{0}}\,H_{0}\ \right]
\ a(t)\ \int_{t_{0}}^{t}\frac{a_{0}}{a^{2}(y)}\,dy \ .
\ee

For the Einstein--de Sitter example, which we referred to before,
$a(t) = a_{0}\,(t/t_{0})^{2/3}$ (as $H_{0} =
\sfrac{2}{3}\,t_{0}^{-1}$), and we find
\be
\eta_{2}(t) = \eta_{2}\,|_{t_{0}}\left(\frac{t}{t_{0}}\right)^{2/3}
+ 3\,\left[\ \left.\frac{d\eta_{2}}{dt}\,\right|_{t_{0}}
- \sfrac{2}{3}\,\eta_{2}\,|_{t_{0}}\,t_{0}^{-1}\ \right]
\,t_{0}^{2/3}\,t^{2/3}\,(\,t_{0}^{-1/3}-t^{-1/3}\,) \ .
\ee

\noindent
Special cases: \enl
{\bf A:} Suppose $\eta_{2}=0$ at $t=t_{0}$ (matter flowing out
isotropically from the central line at that instant), then
\be
\l{etaeds1}
\eta_{2}(t) = 3\,\left.\frac{d\eta_{2}}{dt}\,\right|_{t_{0}}\,
t_{0}^{2/3}\,t^{2/3}\,(\,t_{0}^{-1/3}-t^{-1/3}\,) \ ,
\ee
giving the radial motion of free particles relative to the
fundamental observers, that start off by diverging from them. The
graph of Eq. (\r{etaeds1}) was plotted in Fig \r{fig1}.
\begin{figure*}[!htb]
\vspace*{0.3cm}
\epsfxsize=20pc\epsfbox{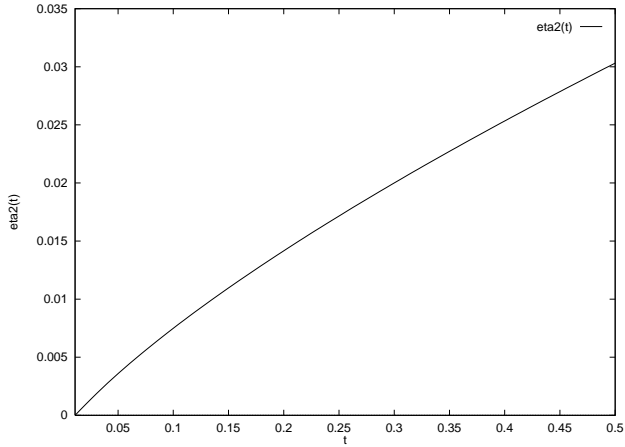}
\vspace*{0.5cm}
\caption{Plot of the deviation vector magnitude $\eta_{2}(t)$
according to Eq. (\r{etaeds1}). The parameter values chosen are
$H_{0} = 60\,km/s/Mpc$, i.e., $t_{0} = 0.01\,(Mpc/km)s$, and
$d\eta_{2}/dt\,|_{t_{0}} = 0.1$.}
\l{fig1}
\end{figure*}
%
\enl

\noindent
{\bf B}: Suppose $d\eta_{2}/dt=0$ at $t=t_{0}$ (matter released
from rest at that instant, hence, not comoving with the expanding
fundamental matter), then
\be
\l{etaeds2}
\eta_{2}(t) = \eta_{2}\,|_{t_{0}}\left(\frac{t}{t_{0}}\right)^{1/3}
\left[\ 2 - \left(\frac{t}{t_{0}}\right)^{1/3}\ \right] \ ,
\ee
gives their radial motion relative to the fundamental
observers. The graph of Eq. (\r{etaeds2}) was plotted in
Fig. \r{fig2}.
\begin{figure*}[!htb]
\vspace*{0.3cm}
\epsfxsize=20pc\epsfbox{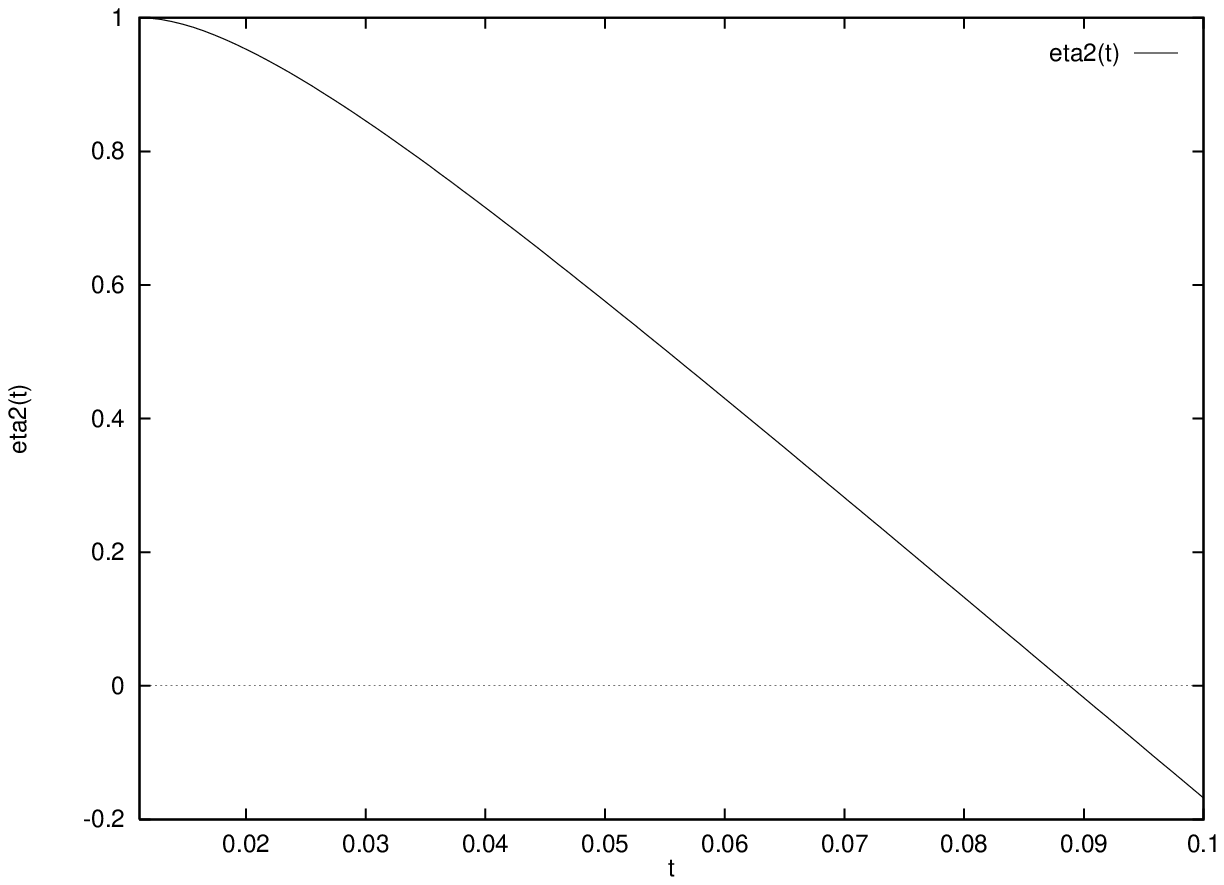}
\vspace*{0.5cm}
\caption{Plot of the deviation vector magnitude $\eta_{2}(t)$
according to Eq. (\r{etaeds2}). The parameter values chosen are
$H_{0} = 60\,km/s/Mpc$, i.e., $t_{0} = 0.01\,(Mpc/km)s$, and
$\eta_{2}\,|_{t_{0}} = 1\,\mbox{unit length}$.}
\l{fig2}
\end{figure*}
\enl

\noindent
{\bf C}: Suppose $d\eta_{2}/dt = \eta_{2}\,|_{t_{0}}\,H_{0}$ at
$t=t_{0}$ (matter initially comoving with the expanding fundamental
matter), then the matter continues to move as the fundamental
observers, i.e., $\eta_{2}(t) = \eta_{2}\,|_{t_{0}}\,
(t/t_{0})^{2/3}$.
\enl

\noindent
Generically, the first integral (\r{gdefi}), applied to this
timelike case, relates the {\em out\/}-going and {\em in\/}-coming
geodesics that link two (timelike separated) points $O$ and $P$, on
fixing boundary conditions for the first integral: namely it
relates the positions and velocities of each congruence at $O$ to
those at $P$. Apart from the cases just considered, the other one
that arises naturally is if particles 1 are at rest at $O$ and
coincide at $P$, whereas for particles 2 the situation is the
converse: they are at rest at $P$ but coincide at $O$. Then
\bea
\eta_{1}\,|_{O}\,\left.\frac{d\eta_{2}}{dt}\,\right|_{O} & = &
\eta_{2}\,|_{P}\,\left.\frac{d\eta_{1}}{dt}\,\right|_{P} \ .
\eea
This relates the positions and velocities at $O$ and $P$, showing
that if both distances are the same (in {\em absolute\/}, not
comoving terms), then the velocities will be the same.

\subsection{Past directed null vector fields}
\l{sec:pastnull}
{\bf Case 2: $V^{a} = k^{a}$, $k_{a}\,k^{a} = 0$, $k^{0} < 0$}.
Equation (\r{devfor2}) now gives
\be
\l{nullfor}
R_{abcd}\,k^{b}\,\eta^{c}\,k^{d} = \sfrac{1}{2}\,(\mu+p)\,E^{2}\,
\eta_{a} \ ,
\ee
so writing $\eta^{a} = \eta\,e^{a}$, $e_{a}\,e^{a}=1$, $0 =
e_{a}\,u^{a} = e_{a}\,k^{a}$, and using a parallelly propagated and
aligned basis, $\delta e^{a}/\delta v = k^{b}\nabla_{b}e^{a}=0$, we
find from (\r{gde}),
\be
\l{nullgde}
\frac{d^{2}\eta}{dv^{2}} = -\,\sfrac{1}{2}\,(\mu+p)\,E^{2}\,\eta \ .
\ee
Again, in line with the timelike case of Eq. (\r{tlgde}), all
families of past-directed (and future-directed) null geodesics
experience focusing, provided $(\mu+p) > 0$ (while the sign of
$\Lambda$ has no influence). Equation (\r{nullgde}) is easily
solved in the case of the de Sitter universe, where $(\mu+p) = 0$,
and the solution is $\eta(v) = C_{1}\,v + C_{2}$, equivalent to the
(flat) Minkowski spacetime case. For null rays diverging from the
origin, $C_{2} = 0$, and we have the same angular size-distance
relation as in flat space (provided we measure distance in terms of
the affine parameter $v$).\enl

When dynamical matter is present, we need to express the quantities
contained in Eq. (\r{nullgde}) in terms of the ({\em non\/}-affine
parameter) redshift $z$, defined in Eq. (\r{red}). A standard
collection of mathematical formulae \ct{brosem87} gives for the
derivative operator of Eq. (\r{nullgde}) the expression
\be
\frac{d^{2}}{dv^{2}} = \left(\frac{dv}{dz}\right)^{-2}\,
\left[\ \frac{d^{2}}{dz^{2}} - \left(\frac{dv}{dz}\right)^{-1}\,
\frac{d^{2}v}{dz^{2}}\,\frac{d}{dz}\ \right] \ .
\ee
{}From Eq. (\r{red}) we know that
\be
(1+z) = \frac{a_{0}}{a} = \frac{E}{E_{0}} \hsp5 \Rightarrow \hsp5
\frac{dz}{(1+z)} = -\,\frac{da}{a} = \frac{dE}{E} \ ,
\ee
hence, (in the past-directed case),
\be
dz = \,(1+z)\,\frac{1}{a}\,\frac{da}{dv}\,dv
= \,(1+z)\,\frac{1}{a}\,\frac{da}{dt}\,E\,dv
= \,E_{0}\,H\,(1+z)^{2}\,dv \ ,
\ee
which leads to
\be
\l{dvdz}
\frac{dv}{dz} = \,\frac{1}{E_{0}\,H\,(1+z)^{2}} \ .
\ee
The Hubble parameter is to be determined via the Friedmann
equation, Eq. (\r{fried}), from which one obtains
\be
\l{hub}
H^{2} = \sfrac{1}{3}\,\mu + \sfrac{1}{3}\,\Lambda
+ H_{0}^{2}\,(1-\Omega_{0}-\Omega_{\Lambda_{0}})\,(1+z)^{2} \ .
\ee
By use of the Raychaudhuri equation, Eq. (\r{tlgde}), one finds,
furthermore,
\be
\frac{d^{2}v}{dz^{2}} = -\,\frac{3}{E_{0}\,H\,(1+z)^{3}}\,
[\ 1 + \frac{1}{18H^{2}}\,(\mu+3p) - \frac{1}{9H^{2}}\,\Lambda\ ]
\ . 
\ee
So, altogether, the null GDE, Eq. (\r{nullgde}), can be expressed
in the new form
\bea
\l{nullgde2}
0 = \frac{d^{2}\eta}{dz^{2}} + \frac{3}{(1+z)}\,[\ 1
+ \frac{1}{18H^{2}}\,(\mu+3p)- \frac{1}{9H^{2}}\,\Lambda\ ]
\ \frac{d\eta}{dz}
+ \frac{1}{2(1+z)^{2}}\,\frac{1}{H^{2}}\,(\mu+p)\ \eta \ .
\eea

If we consider again the non-interacting mixture of incoherent
matter and radiation, we have
\be
\mu = 3\,H_{0}^{2}\,\Omega_{m_{0}}(1+z)^{3}
+ 3\,H_{0}^{2}\,\Omega_{r_{0}}(1+z)^{4} \ , \hsp5
p = H_{0}^{2}\,\Omega_{r_{0}}(1+z)^{4} \ .
\ee
Then, from Eq. (\r{hub}), for $\Lambda=0$ the Hubble parameter
evaluates to
\be
\l{hub2}
H^{2} = H_{0}^{2}\,(\ 1 + \Omega_{m_{0}}\,z
+ \Omega_{r_{0}}\,z\,(2+z)\ )\,(1+z)^{2} \ ,
\ee
and Eq. (\r{nullgde2}) assumes the form
\bea
\l{nullgde3}
0 & = & \frac{d^{2}\eta}{dz^{2}}
+ \frac{6+\Omega_{m_{0}}(1+7z) + 2\Omega_{r_{0}}(1+8z+4z^{2})}{
2\,(1+z)\,(\ 1+\Omega_{m_{0}}z + \Omega_{r_{0}}z(2+z)\ )}
\ \frac{d\eta}{dz} 
\nonumber \\ & & \hsp5 
+ \ \frac{3\Omega_{m_{0}}
+ 4\Omega_{r_{0}}(1+z)}{2\,(1+z)\,(\ 1 + \Omega_{m_{0}}z
+ \Omega_{r_{0}}z(2+z)\ )}\ \eta \ .
\eea
When only incoherent matter is present (the dust case), then
$\Omega_{r_{0}} = 0$, while a sole incoherent radiation matter
source has $\Omega_{m_{0}} = 0$. The popular spatially flat FLRW
case is contained for $\Omega_{0} = \Omega_{m_{0}} + \Omega_{r_{0}}
= 1$.

The general solution to this linear homogeneous second-order
ordinary differential equation is given by
\bea
\l{nullgdesol}
\eta(z) = \frac{1}{(1+z)^{2}}\,[\ C_{1}\,(\,2 - \Omega_{m_{0}}
- 2\Omega_{r_{0}} + \Omega_{m_{0}}z\,)
+ C_{2}\,(\,1 + \Omega_{m_{0}}z + \Omega_{r_{0}}z(2+z)\,)^{1/2}\ ]
\ ,
\eea
which we obtained with support from some computer algebra
packages. The integration constants $C_{1}$ and $C_{2}$ carry the
dimension of $\eta(z)$.  With this explicit form for the deviation
vector of a (past-directed) geodesic null congruence at our hands,
we are, now, in a position to easily infer an expression for the
observer area distance, $r_{0}(z)$, originally derived by Mattig
\ct{mat58} for the dust case ($\Omega_{r_{0}}=0$), which is of
considerable astronomical importance (\,see, e.g., Refs. \ct{san61}
and \ct{ell71}\,). Using $d/d\ell = E_{0}^{-1}\,(1+z)^{-1}\,d/dv =
H\,(1+z)\,d/dz$ (\,cf. Eqs. (\r{dldv}) and (\r{dvdz})\,) and
choosing the integration constants in Eq. (\r{nullgdesol}) such
that $\eta(z=0) = 0$, its definition\footnote{$d\Omega_{0}$ here
denotes an infinitesimal solid angle rather than a change in
density parameter.},
$$
r_{0}(z) := \sqrt{\ \left|\,\frac{dA_{0}(z)}{d\Omega_{0}}\,
\right|\ } = \left|\ \frac{\left.\eta(z')\,\right|_{z'=z}}{\left.
d\eta(z')/d\ell\,\right|_{z'=0}}\ \right| \ ,
$$
yields 
\bea
\l{oad}
r_{0}(z) & = & H_{0}^{-1}\,[\ 2\Omega_{m_{0}}
- (\,2 - \Omega_{m_{0}} - 2\Omega_{r_{0}}\,)\,(\,\Omega_{m_{0}}
+ 2\Omega_{r_{0}}\,)\ ]^{-1} \nonumber \\
& & \hspace{1cm} \times \frac{2}{(1+z)^{2}}\,[\ (\,2
- \Omega_{m_{0}} - 2\Omega_{r_{0}} + \Omega_{m_{0}}z\,) \\ 
& & \hspace{3cm} - \ (\,2 - \Omega_{m_{0}}
- 2\Omega_{r_{0}}\,)\,(\,1 + \Omega_{m_{0}}z
+ \Omega_{r_{0}}z(2+z)\,)^{1/2}\ ] \ , \nonumber
\eea
giving the observer area distance as a function of the redshift $z$
in units of the present-day Hubble radius $H_{0}^{-1}$ for an
arbitrary non-interacting mixture of matter and radiation (\,and
containing as a special case the Mattig formula when $\Omega_{r_0}
=0$\,). The graph of Eq. (\r{oad}) was plotted in Fig. \r{fig3}.

The formula (\r{oad}) is equivalent to the one stated earlier by
Matravers and Aziz \ct{matazi88}, but --- unlike the usual
calculations --- is obtained in a uniform way from the null GDE
(\,irrespective of the intrinsic curvature of the spacelike
3-surfaces $\{t = \mbox{const}\}$\,). In the usual approach, three
separate calculations are needed (one for each value of $k$), and
it is a matter of some amazement that they all fit the same formula
in the end. In the present approach, {\em one\/} integration is
needed, leading to one formula --- a considerable increase in
clarity.
\begin{figure*}[!htb]
\vspace*{0.3cm}
\epsfxsize=20pc\epsfbox{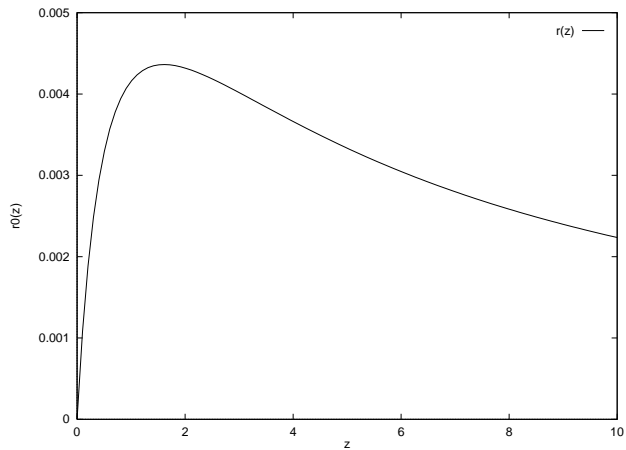}
\vspace*{0.5cm}
\caption{Plot of the observer area distance $r_{0}(z)$ according to
Eq. (\r{oad}), in units of $H_{0}^{-1}$. The parameter values
chosen are $H_{0} = 80\,km/s/Mpc$, $\Omega_{m_{0}} = 0.2$ and
$\Omega_{r_{0}} = 0.1$.}
\l{fig3}
\end{figure*}

The first integral relation can be investigated analogously to the
timelike case above. Consider null rays diverging from the observer
at $O$ and arriving at the source $S$, with deviation vector
$\eta_1$, and null rays diverging from the source $S$ and arriving
at the observer $O$, with deviation vector $\eta_{2}$. The first
integral is the same as before, but now we need to convert (for
past-directed null rays) from the affine parameter $v$ to $\ell$
according to Eq. (\r{dldv}), with $a_{0}/a=(1+z)$. One obtains
\bea
\l{nrt}
\eta_{2}\,|_{O}\,\left.\frac{d\eta_{1}}{d\ell}\,\right|_{O} & = &
\eta_{1}\,|_{S}\,\left.\frac{d\eta_{2}}{d\ell}\,\right|_{S} (1+z)
\ , 
\eea
where the terms $d\eta/d\ell$ are the angles subtended by the pairs
of null rays corresponding to the deviation vectors. Expressed in
terms of angular diameter distances, $r_{O}$ and $r_{S}$, defined
by
\be
\eta_{1}\,|_{S} := r_{O}\,\left.\frac{d\eta_{1}}{d\ell}\,
\right|_{O} \ , \hsp5
\eta_{2}\,|_{O} := r_{S}\,\left.\frac{d\eta_{2}}{d\ell}\,
\right|_{S} \ ,
\ee
(which, for FLRW geometry, are the {\em same\/} as area distances), 
we find the familiar null reciprocity theorem for FLRW models 
\ct{wei72,ell87}:
\be
r_{S} = r_{O}\,(1+z) \ .
\ee
This underlies the equivalence (up to redshift factors) of area
distance and luminosity distance, and the fact that measured
radiation intensity is independent of area distance, depending only
on redshift (\,see Ref. \ct{ell71} for a more detailed
discussion\,). These features are fundamental in analysing
observations of distant sources and measurements of the cosmic
microwave background radiation.

\subsection{Generic geodesic vector fields}
\l{sec:gengeo}
{\bf Case 3:} $V_{a}\,V^{a} = \eps$, {\em not\/} parallel to
$u^{a}$, {\em nor\/} null. The force term in the generic case is
provided by Eq. (\r{devfor2}). Writing $\eta^{a} = \ell\,e^{a}$,
$e_{a}\,e^{a}=1$, $0 = e_{a}\,V^{a} = e_{a}\,u^{a}$, and employing
a parallelly propagated and aligned basis, $\delta e^{a}/\delta v =
V^{b}\nabla_{b}e^{a} = 0$, we find from Eq. (\r{gde}),
\be
\l{gengde}
\frac{d^{2}\ell}{dv^{2}} = -\,\eps\,\sfrac{1}{3}\,(\mu+\Lambda)
\,\ell - \sfrac{1}{2}\,(\mu+p)\,E^{2}\,\ell \ ,
\ee
giving the spatial orthogonal separation of these geodesics within
the 2-D screen space as they spread out in spacetime.

\subsubsection{Orthogonal spacelike geodesics}
A particular case is the spatial geodesics that start off {\em
orthogonal\/} to $u^{a}$ (so $E_{0} = 0$, which implies that the
corresponding geodesics are indeed spacelike), but then bend down
towards the past thereafter (see the discussion in section
\r{sec:slgeod}). The above equation applies with $\epsilon = 1$.
The simplest case is a de Sitter universe where $0 = \mu = p$, and
then the solution for {\em all\/} $|\,E_{0}\,| \geq 0$, i.e., {\em
all\/} spacelike geodesics is
\bea
\ell(v) = \left\{ \begin{array}{ll}
          C_{1}\,\cos(\alpha\,v) + C_{2}\,\sin(\alpha\,v)
          & \Lambda > 0 \\
          C_{1}\,\cosh(\alpha\,v) + C_{2}\,\sinh(\alpha\,v)
          & \Lambda < 0 \\
                  \end{array} \right. \ ,
\eea
with $\alpha := (\sfrac{1}{3}\,|\,\Lambda\,|)^{1/2}$ (note this is
just the exact converse to the timelike case of Eq. (\r{tldssol})
above).

In the case of non-zero dynamical matter, however, $\mu$ and $p$
are {\em not\/} constants along the initially orthogonal geodesics,
as these geodesics do {\em not\/} remain within a spacelike
3-surface $\{t = \mbox{const}\}$; we have to find $\mu[\,t(v)\,]$
or $\mu[\,t(r)\,]$ from the geodesic equation. However, near the
starting point $v=0$ at $t_{0}$ we have (for $\Lambda = 0$)
\be
0 = \left.\frac{d^{2}\ell}{dv^{2}}\,\right|_{t_{0}}
+ \sfrac{1}{3}\,\mu_{0}\,\ell \ ,
\ee
giving the solution near this origin, on carrying out a first-order
expansion, by
\be
\ell(v) = \ell_{0}\,\cos(\omega_0\,v) + \left.\frac{d\ell}{dv}\,
\right|_{t_{0}}\,\sin(\omega_0\,v) \ , \hsp5
\omega_{0} := (\sfrac{1}{3}\,\mu_{0})^{1/2} \ .
\ee
This is always convergent for normal matter, irrespective of the
intrinsic curvature of the particular spacelike 3-surface $\{t_{0}
= \mbox{const}\}$ considered. However, as soon as the distance is
appreciable, the geodesics will have bent down and lie below the
initial 3-surface $\{t_{0} = \mbox{const}\}$, where the density of
matter will be higher and the curvature greater. Thus, the
geodesics will tend to converge even more strongly.

\subsubsection{Geodesics in the orthogonal spacelike 3-surfaces}
This is to be contrasted with geodesic congruences {\em within\/}
the spacelike 3-surfaces $\{t = \mbox{const}\}$ orthogonal to
$u^{a}$, which are 3-spaces of maximal symmetry. In contrast to
Eq. (\r{geod}), these geodesics satisfy the 3-D equations
\be
\l{3dgeod}
V^{a} := \frac{dx^{a}(v)}{dv}\ , \hsp5 V_{a}\,V^{a} = 1 \ ,
\hsp5 V_{a}\,u^{a}=0, \hsp5 0 = V^{b}\D_{b}V^{a} \ .
\ee
{}From Eq. (\r{3d}), the force term for the resulting 3-D spatial
GDE\footnote{Determined by Eqs. (\r{com}), (\r{orth}), and
(\r{3dgeod}).} takes the form
\be
{}^{3}\!R_{abcd}\,V^{b}\,\eta^{c}\,V^{d} = \sfrac{1}{3}\,(\,\mu
-\sfrac{1}{3}\,\Th^{2}+\Lambda\,) \,\eta_{a} = K\,\eta^{a} \ ,
\ee
where $K(t)$ is the curvature of these 3-spaces (\,cf. Eq.
(\r{eq:gauss})\,).  Consequently, whether geodesics in these
spacelike sections converge or diverge depends on the sign of
$K$. Setting $\eta^{a} = \eta\,e^{a}$ where $e_{a}\,e^{a}=1$ and
$e_{a}\,u^{a}=0$, as before we choose a congruence of vectors such
that $\delta e^{a}/\delta v = V^{b}\D_{b}e^{a} = 0$ and the 3-D
spatial GDE\footnote{That is, the 3-D version of Eq. (\r{gde}) that
applies in these 3-spaces.} becomes
\be
\l{3dgde}
\frac{d^{2}\eta}{d v^{2}} = -\,K\,\eta \ .
\ee
$K=K(t)$ is indeed {\em constant\/} along these spatial geodesics
(because they lie within the 3-surfaces $\{t = \mbox{const}\}$). If
$K > 0$, one deals again with the familiar oscillator equation,
i.e., two neighbouring spatial geodesics will harmonically converge
to and diverge from each other as $v$ increases. If $K < 0$, they
will exponentially diverge, and if $K = 0$, they diverge linearly.
\enl

Focusing on radial spatial geodesics, the local FLRW coordinates of
the spacelike 3-surfaces $\{t = \mbox{const}\}$ arise as
follows. We consider a 3-space with metric
$f_{\mu\nu}\,dx^{\mu}\,dx^{\nu}$, and constant dimensionless scalar
curvature, if non-zero, normalised to $k = \pm\,1$,
(\,cf. Eqs. (\r{ds2}) and (\r{3curvscl})\,). Note that the full
3-space metric $h_{\mu\nu}(t)$ at arbitrary time $t$ is just given
by $h_{\mu\nu}(t) = a^{2}(t)\,f_{\mu\nu}$.\footnote{When $a(t)$ is
of unit magnitude, say at time $t = \tilde{t}$, then $f_{\mu\nu}$
is equal to the metric $h_{\mu\nu}(\tilde{t})$ on the 3-surface
$\{\tilde{t} = \mbox{const}\}$, except for a dimensional unit
factor, and similarly for $k$ and $K(\tilde{t})$.} Choosing an
affine parameter $v=r$, $V^{a} = (\partial_{r})^{a} =
\delta^{a}\!_{1}$ is the geodesic unit normal to the 2-surfaces $\{r =
\mbox{const}\}$, which are 2-spheres of area
$4\pi\,f^{2}(r)$. Thus, it is tangent to the orthogonal coordinate
curves $x^{2} = \mbox{const}$, $x^{3} = \mbox{const}$. A basis of
deviation vectors in the 2-D screen space is given by
$\eta_{1}{}^{a} = \delta^{a}\!_{2}$ and $\eta_{2}{}^{a} =
\delta^{a}\!_{3}$ (these commute with the geodesic vector $V^{a} =
\delta^{a}\!_{1}$, because each of these is a coordinate basis
vector). Employing an {\em orthonormal\/} basis with components
$(e_{1})^{a} = \delta^{a}\!_{1}$, $(e_{2})^{a} =
f^{-1}(r)\,\delta^{a}\!_{2}$, $(e_{3})^{a} =
f^{-1}(r)\,(\sin\theta)^{-1}\,\delta^{a}\!_{3}$, parallelly
propagated along the radial geodesics $V^{a}$, Eq. (\r{3dgde})
yields
\be
\l{4dgde}
0 = \frac{d^{2}\eta}{d v^{2}} + k\,\eta \hsp5 \Rightarrow
\hsp5
0 = \frac{d^{2}f}{dr^{2}} + k\,f \ ;
\ee
the second relation following because relative to the orthonormal
basis, $\eta_{1}{}^{a} = f(r)\,\delta^{a}\!_{2}$ and
$\eta_{2}{}^{a} = f(r)\,\sin\theta\,\delta^{a}\!_{3}$ (apply the
first equation to either vector to get the second). Then the
solution we want corresponds to {\em that\/} solution for which
$\eta(r=0) = 0$; we find
\bea
f(r) = \left\{ \begin{array}{rl}
       \sin\,r  & k = +\,1 \\
       r        & k = 0 \\
       \sinh\,r & k = -\,1
               \end{array} \right. \ ,
\eea
showing how the GDE within the spacelike 3-surfaces $\{t =
\mbox{const}\}$ determines the function $f(r)$ in
Eq. (\r{ds2}). The corresponding solutions with $d\eta/dr = 0$ at
$r = 0$ exhibit precisely how Euclid's parallel postulate breaks
down for these curved 3-space sections, according to the spatial
curvature.

In this context it is of interest to remark that the
Lorentz-invariant\footnote{That is, $u^{a}$ is {\em not} uniquely
defined.} de Sitter spacetime geometry, which is the case $0 = \mu
= p$, $\Lambda > 0$, can be sliced by spacelike 3-surfaces $\{t =
\mbox{const}\}$ of either constant positive, zero, or negative
intrinsic curvature (\,cf. Ref. \ct{sch56}\,), depending on the
sign of the sum $3\,K = -\,\sfrac{1}{3}\,\Th^{2} + \Lambda$ (\,see
Eq. (\r{eq:gauss})\,). For anti-de Sitter ($\Lambda < 0$) only the
negative curvature case applies. The different FLRW forms of the de
Sitter spacetime metric follow from arguments essentially identical
to that just given for the 3-space metric, because it is a 4-space
of constant curvature, i.e., maximal symmetry (\,and the argument
applies also to the 2-sphere, leading to the form of the terms in
the last bracket in Eq.  (\r{ds2})\,). In each case, the GDE,
together with the constant curvature condition (\r{cc}), leads to
the harmonic equation (\r{4dgde}).  \enl

Similarly to the null case, the 3-D geometrical reciprocity theorem
can be stated as
\bea
\l{3drt}
\eta_{1}\,|_{O}\,\left.\frac{d\eta_{2}}{dr}\,\right|_{O}
& = &
\eta_{2}\,|_{P}\,\left.\frac{d\eta_{1}}{dr}\,\right|_{P} \ ,
\eea
showing how geodesics diverging about a central geodesic from $P$
to $O$ at an angle $\alpha_0$ reach a separation $d$ at $O$, and
geodesics diverging from $O$ at the same angle will reach the same
distance apart at $P$ (irrespective of the spatial curvature which
is constant). Corresponding statements hold for the families of
geodesics that diverge from $P$ and $O$, and end up parallel at $O$
and $P$, respectively.

\section{Conclusion}
One way of solving the EFE is to treat them as {\it algebraic}
equations relating $R_{abcd}$ to $R_{ab}$ and $C_{abcd}$, then
solving the GDE (which characterises relative acceleration due to
spacetime curvature) to determine both the spacetime geometry and
its properties. In the case of a FLRW model, this can be carried
out explicitly, as shown above: integrating the GDE
(\,cf. Eqs. (\r{tlgde}), (\r{nullgde2}) and (\r{gengde})\,) allows
complete characterisation of all interesting geometrical features
of the exact FLRW geometry in an elegant manner --- determining the
timelike evolution, spacelike geometry, and null ray properties,
which in turn determine the basic observational properties.  The
Newtonian analogue of some of this has been given by Tipler
\ct{tip96a,tip96b}.

An interesting project is to extend this calculation to perturbed
FLRW models in order to work out the effects of linear anisotropies
on the present results as regards all three causal cases (timelike,
spacelike, null). This would allow investigation of both dynamical
and observational features of such models, for example examining
aspects of gravitational lensing theory \ct{schetal92}.\enl

\section*{Acknowledgements}
This work has been supported by the South African Foundation for
Research and Development (FRD). The integration of the GDEs was
facilitated by application of the computer algebra packages {\tt
MAPLE} and {\tt REDUCE}.



\end{document}